\documentclass[12pt]{article}
\usepackage{amsmath}
\usepackage{amssymb}
\begin{document}

\vskip 2cm
\begin{center}
{\bf {\Large  Canonical brackets of a toy model for the Hodge theory\\
without its canonical conjugate momenta}}\\

\vskip 3.2cm

{\sf D. Shukla$^{(a)}$, T. Bhanja$^{(a)}$, R. P. Malik$^{(a,b)}$}\\
$^{(a)}$ {\it Physics Department, Centre of Advanced Studies,}\\
{\it Banaras Hindu University, Varanasi - 221 005, (U.P.), India}\\

\vskip 0.1cm


\vskip 0.1cm

$^{(b)}$ {\it DST Centre for Interdisciplinary Mathematical Sciences,}\\
{\it Faculty of Science, Banaras Hindu University, Varanasi - 221 005, India}\\
{\small {\sf {E-mails: dheerajkumarshukla@gmail.com; tapobroto.bhanja@gmail.com; 
rpmalik1995@gmail.com}}}

\end{center}

\vskip 3cm

\noindent
{\bf Abstract:}We consider the toy model of a rigid rotor as an example of the Hodge theory 
within the framework of Becchi-Rouet-Stora-Tyutin (BRST) formalism 
and show that the {\it internal} symmetries of this theory lead to the derivation 
of canonical brackets amongst the creation and annihilation operators 
of the dynamical variables where the definition of the 
canonical conjugate momenta is {\it not} required. We invoke {\it only} 
the spin-statistics theorem, normal ordering and basic concepts of 
continuous symmetries (and their generators) to derive the canonical
brackets for the model of a one (0 + 1)-dimensional (1D) rigid rotor without 
using the definition of the canonical conjugate momenta {\it anywhere}. Our present 
method of derivation of the basic brackets is conjectured to be true 
for a class of theories that provide a set of tractable physical examples 
for the Hodge theory.\\

\vskip 1.2cm



\newpage
\noindent
\section{Introduction}

One of the earliest methods of quantization of a classical (physical)
system is the standard canonical quantization scheme where the 
(graded)Poisson brackets of the classical mechanics are upgraded  
to the (anti)commutators at the quantum level. In this theoretical
set-up, we invoke primarily {\it three} basic ideas. First, we 
distinguish between the fermionic and bosonic variables by invoking the 
idea of spin-statistics theorem. Second, we take the help of the definition 
of canonical conjugate momenta to obtain the momenta corresponding 
to all the dynamical variables of a given classical theory and define 
the (graded)Poisson brackets. These brackets are then elevated to the 
(anti)commutators between the variables and corresponding momenta in 
their operator form. If the equations of the motion of the theory support 
the existence of creation and annihilation operators, the above canonical 
(anti)commutators are translated into the basic (anti)commutators 
amongst the creation and annihilation operators (e.g. in the problem 
of simple harmonic oscillator of quantum mechanics) and the 
quantization follows (at the algebraic level amongst the creation and 
annihilation operators). Finally, to make the physical sense 
out of some of the important quantities like Hamiltonian, conserved 
charges, etc., it is essential to adopt the normal ordering 
procedure in which  the creation operators are brought to the 
left in all the terms that are found to be present in the above mentioned
physical quantities of interest in a given theory.

One can provide physical meaning to the concepts of spin-statistics 
theorem and normal ordering but the definition of the canonical 
conjugate momenta remains mathematical in nature. In our present 
endeavor, we demonstrate that one can perform the canonical quantization 
without taking the help of the definition of canonical conjugate 
momenta for a {\it class} of theories which are models for the {\it Hodge 
theory}. The latter models are physical examples where the symmetries of 
the theory provide the physical realizations of the de Rham cohomological 
operators\footnote{On a compact manifold without a boundary, a set of 
{\it three} operators ($d, \delta, \Delta$) is called the de Rham 
cohomological operators where d (with $d^2 = 0$) is the exterior derivative, 
$\delta = \pm * d \, * $ (with ${\delta}^2 = 0$) is the co-exterior derivative 
and $\Delta$ is the Laplacian operator which obey {\it together} the algebra: 
$[\Delta,\,d] = [\Delta,\,\delta] = 0,$ $\, d^2 = {\delta}^2 = 0,\,
\Delta = (d + \delta)^2 = \{d,\,\delta \}$. In the above, the ($*$) operator 
is popularly known as the Hodge duality operation on a given manifold 
(see, e.g. [1-5] for details) and this algebra is known as Hodge
algebra where $\Delta$ behaves like the Casimir operator (but {\it not} in the 
sense of the Casimir operators of the Lie algebras).} of differential geometry 
[1-5]. To be precise, in our present investigation, we take up a toy model for 
a rigid rotor (which is a model for the Hodge theory [6]) to demonstrate that 
one can quantize this theory without taking the help of canonical conjugate 
momenta. In fact, we exploit the idea of symmetry principles (i.e. continuous 
symmetries and their generators) to obtain the canonical basic brackets which 
are consistent  with the standard canonical method of quantization for this 
system at the level of creation and annihilation operators.

It is crystal clear, from the above assertion, that we shall take 
the help of spin-statistics theorem\footnote{For the one (0 + 1)-dimensional 
toy model, there is no meaning of spin. However, in our present investigation, 
we interpret the spin-statistics theorem in the language of the (anti)commutation 
relations of the dynamical variables of our theory.} as well as normal ordering 
in our present endeavor but we shall {\it not} use canonical conjugate 
momenta {\it anywhere}. This exercise, in some sense, 
provides the physical meaning to the canonical conjugate momenta in the 
language of symmetry principles. Thus, the main result of our present investigation 
is the theoretical trick, we have developed over  the years [7, 8], by which, 
we obtain the basic brackets for the model of the rigid rotor by exploiting the 
symmetry principles (instead of using canonical conjugate momenta) that are 
consistent (and in complete agreement) with the canonical quantization 
scheme\footnote{ It is obvious that we have already exploited our present idea in the 
quantization of 2D {\it free} as well as {\it interacting} Abelian 1-form gauge 
theory [7, 8]. In the latter category, we have considered the topic of QED with 
Dirac fields (where there is a coupling between the photon and a system of charged fermionic
particles).}.

In our present investigation, we have exploited \textit{six} continuous symmetry 
transformations 
to obtain the canonical brackets that are in full agreement with the (anti)commutators 
obtained by using the {\it standard} canonical method of quantization. The key point, 
to be noted, is that {\it all} the six continuous symmetries and their generators play 
important roles in the derivation of {\it all} the possible (non-)vanishing brackets 
that are allowed amongst six creation and six annihilation operators that are present 
in the normal mode expansions (see, (18) below) of the six variables of the first order 
Lagrangian (2) (see below). Thus, we observe that, for the 1D rigid rotor, all the 
continuous symmetries {\it together} play very crucial role in the derivation of 
{\it all} the appropriate (anti)commutators amongst the creation and annihilation
operators at the quantum level.

Our present investigation is essential on the following counts. First and 
foremost, it is very important for us to put our ideas of previous works 
[7, 8] on firmer footings by applying those ideas to some new physical systems 
so that we could get an alternative to the canonical method of quantization for 
a specific class of models that are physical examples of the Hodge theory. Our 
present endeavor is an attempt in that direction. Second, it is always 
gratifying to replace some mathematical definitions by a few physical principles. 
In our present investigation, we have an alternative to the definition 
of canonical conjugate momenta in the sense that we replace it by the symmetry 
principles for the quantization of our present system. Third, our method of 
quantization adds richness and variety in theoretical physics even though it 
is applied to a special class of theories that are examples of the Hodge theory. 
Finally, our present endeavor is a part of our first few steps towards our main 
goal of the proof that, for the models of the Hodge theory, the definition of 
canonical conjugate momentum is {\it not} required as far as the quantization 
of these models is concerned within the framework of BRST 
formalism\footnote{We have also shown that the ${\mathcal N} = 2$ SUSY quantum mechanical 
models are also a set of examples for the Hodge theory which are not discussed 
within the framework of BRST approach (cf. Sec. 8 below).}.

The material of our present investigation is organized as follows. 
We discuss the continuous symmetries and derive the corresponding 
Noether conserved charges in our Sec. 2. In our forthcoming Sec. 3,
we describe the {\it standard} canonical quantization of a 1D model 
for the rigid rotor. Sec. 4 contains the derivation of basic brackets 
from the ghost symmetry transformations where we do not use the 
definition of canonical conjugate momenta. Our Sec. 5 is devoted 
to the derivation of (anti)commutators from the basic symmetry 
principles associated with the continuous (anti-)BRST symmetry 
transformations. We derive the (anti)commutators by taking the 
help of basic concepts of (anti-)co-BRST symmetry transformations 
and their Noether conserved charges in Sec. 6. Our Sec. 7 contains 
the derivation of the same brackets from the bosonic symmetry 
transformations. Finally, we make some concluding remarks in Sec. 8
and point out a few future directions.

In our Appendix A, we have obtained the explicit canonical basic
brackets from the {\it standard} canonical quantization method for 
the sake of precise comparison with such kind of brackets derived 
in the main body of our text. Our Appendix B is devoted to some 
comments on the mode expansions that have been quoted in Eq. (18)
(cf. Sec. 3) of our present endeavor.

\label{Notations}
{\it General Notations and Convention}: Throughout the whole body of 
our text, we denote the (anti-)BRST and (anti-)dual-BRST [i.e.(anti-)
co-BRST] symmetry transformations by $s_{(a)b}$ and $s_{(a)d}$, respectively. 
Various forms of the Lagrangians (that respect the above symmetries) have 
been denoted with a subscript $(B)$ attached to them. Furthermore, we have 
adopted the convention of left-derivative w.r.t. fermionic variables of 
our theory everywhere in our present endeavor.

\noindent
\section{Preliminaries: Symmetries and Charges}

We begin with the (anti-)BRST invariant first order Lagrangian 
(see e.g. [9, 6, 10]) for the rigid rotor (with mass 
$m = 1$) as follows:
\begin{eqnarray}
L_0 = \dot r \, p_r + \dot\theta \, p_{\theta} - \frac{p_{\theta}^2}{2\,r^2} 
- \lambda \,(r - a) + B\,(\dot \lambda - p_r)  + \frac{1}{2}\,B^2
- i\, \dot{\bar C}\,\dot C + i\, \bar C\, C,
\end{eqnarray}
where ($r, \theta $) are the polar coordinates, ($p_r, p_\theta$) are 
the corresponding conjugate momenta, $\lambda$ is the ``gauge" variable, 
$ B $ is the Nakanishi-Lautrup type auxiliary variable and 
$ (\bar{C})C $ are the  fermionic $ (C^{2} = 0 = {\bar{C}}^2,\; 
C\,\bar{C} + \bar{C}\,C = 0)$ (anti-)ghost variables. Here 
$\dot{\lambda} = d\,\lambda/dt,\; \dot{r} = dr/dt,\; 
\dot{\theta} = d\theta/dt $, etc., are the generalized ``velocities"  
of the dynamical variables  with respect to the evolution parameter 
$t$ of our theory. The auxiliary variable $B$ is invoked to linearize 
the gauge-fixing term $\bigl[-\,(\dot{\lambda} - p_r)^2 /2\bigr]$ which 
contains $\dot{\lambda}$ and $p_r$ together. There are two first-class 
constraints on the theory which originate from $(r - a) \approx 0 $ 
and $d/dt\,(r - a) \approx 0 $ (where $a$ is the radius of the circle 
on which a particle of unit mass $(m = 1)$ moves in the system of a rigid rotor). 
We can get rid of one of the auxiliary variables by using the 
Euler-Lagrange (EL) equations of motion (e.g. $ p_\theta = r^2\, \dot{\theta}$). 
The ensuing Lagrangian   
\begin{eqnarray}
L_B = \dot r \, p_r + \frac{1}{2}\,r^{2}\,{\dot\theta}^{2} 
- \lambda \,(r - a) + B\,(\dot{\lambda} - p_r) + \frac{1}{2}\,B^2
- i\, \dot{\bar C}\,\dot C + i\, \bar C\, C,
\end{eqnarray}
respects the following off-shell nilpotent $ (s_{(a)b}^2 = 0) $ 
continuous (anti-)BRST symmetry transformations  $ (s_{(a)b})$ 
(see e.g. [9, 10, 6] for details):
\begin{eqnarray}
s_b\, p_r &=& - \,C,\qquad\; s_b\, \lambda = \dot C,\, \,\,\qquad \, 
s_b \, \bar C = +\,i\, B, \qquad s_b \,[r, \theta, C, B] = 0,\nonumber\\
s_{ab}\, p_r &=& - \,\bar C,\qquad s_{ab}\, \lambda = \dot{\bar C}, 
\qquad  s_{ab}\, C =  -\,i \,B, \qquad s_{ab}\, [r, \theta, \bar C, B] = 0.
\end{eqnarray}
It is trivial to note that the off-shell nilpotency $ (s_{(a)b}^2 = 0) $ and 
absolute anticommutativity $(s_b\,s_{ab} + s_{ab}\,s_{b} = 0) $ properties 
are {\it true} for the above transformations  $s_{(a)b} $.
Under the  continuous symmetry transformations (3), the 
Lagrangian (2) of our theory transforms to the total time derivatives as:
\begin{eqnarray}
s_b\, L_B =  \frac{d}{dt}\,\bigl [B\,\dot C  - (r-a)\,C \bigr], \qquad\quad 
s_{ab}\, L_B =  \frac{d}{dt}\,\bigl [ B\,\dot {\bar C} - (r-a)\,\bar C \bigr].
\end{eqnarray}
Thus, the transformations (3) are the {\it symmetry} transformations 
for the action integral ($S = \int {dt\, L_B}$). The Noether charges 
(that emerge from the transformations (3)) are as follows:
\begin{eqnarray}
Q_b = B\,\dot{C} - \dot{B}\,C,  \qquad\qquad 
Q_{ab} &=& B\,\dot{\bar C} - \dot{B}\,\bar C.
\end{eqnarray}
The conservation of the charges (according to Noether's theorem) can 
be proven by exploiting the following EL equations of motion (EOM)
\begin{eqnarray}
&& \dot{p_{r}} + \lambda = r\,\dot{\theta}^2, \qquad\quad  
 \dot{B} + (r - a) = 0, \qquad  B + (\dot{\lambda} 
- p_r) = 0, \nonumber\\
&& B = \dot{r} \Rightarrow B = \frac{d}{dt}\,(r - a),\qquad \ddot{C} 
+ C = 0, \qquad\quad\; \ddot{\bar{C}} + \bar{C} = 0,
\end{eqnarray}
which emerge from the Lagrangian (2). It is clear that the physicality 
condition with the (anti-)BRST charges $ Q_{(a)b}\mid phys> = 0 $ 
implies that $ (r - a) \mid phys> = 0 $ and $ (\dot \lambda - p_r)\mid phys> 
= 0 $. Translated in terms of the auxiliary variable $B$, these conditions imply 
that $B \mid phys> = 0$ and $\dot{B} \mid phys> = 0$.
Using the above equations of motion (6), we observe that 
$ (\dot\lambda - p_r )\mid phys> = 0 $ is equivalent to 
$ d/dt \, (r - a) \mid phys> = 0 $. Physically, 
these conditions imply that the motion of the particle is confined 
to a circle of radius $a$ (i.e. $r = a) $ and it remains time-evolution 
invariant (i.e. $d/dt\,(r - a) = 0 $). We note, in passing, that the 
above equations of motion imply that $\ddot{B} + B = 0, 
\quad \frac{d^2}{dt^2}\,(\dot{\lambda} - p_r) 
+ (\dot{\lambda} - p_r) = 0$ and $\ddot{R} + R = 0$
if we identify $R$ with $(r - a)$ (i.e. $R = (r - a)$).
With this identification, the conserved (anti-)BRST charges (5) 
can be re-expressed as: $ Q_b = R\,C + \dot{R}\,\dot{C}, \quad  Q_{ab} 
=  R\,\bar{C} + \dot{R}\,\dot{\bar{C}}$.

We observe that the Lagrangian (2) respects another set of 
nilpotent $ (s_{(a)d}^2 = 0) $ and absolutely anticommuting 
$(s_d\,s_{ad} + s_{ad}\,s_{d} = 0) $ (anti-)co-BRST symmetry 
transformations $ s_{(a)d} $. These transformations are as follows
(see, e.g. [6, 10]):
\begin{eqnarray}
s_d\, \lambda &=& \bar C, \qquad \;  s_d\, C = i\, (r-a), \,\quad\qquad  
s_d\, p_r = \dot{\bar C},\qquad\;\,  s_d \,[B, \bar C, r, \theta] = 0,\nonumber\\
s_{ad}\, \lambda &=& C, \qquad s_{ad}\, \bar C = - i\, (r - a), 
\qquad  s_{ad}\, p_r =  \dot C,\qquad  s_{ad}\, [B,  C, r, \theta] = 0.
\end{eqnarray}
It is elementary to check that $ s_{(a)d}\, L_B  = 0$. We note that\footnote{The 
total gauge-fixing term remains invariant under the (anti-)co-BRST symmetry 
transformations $s_{(a)d}$. This is a characteristic feature of the nilpotent 
(anti-)co-BRST [(anti-)dual-BRST] symmetry transformations $s_{(a)d}$ for this 1D 
system of Hodge theory [6]. We have adopted the notation $(s_{(a)d})$ for the 
infinitesimal and continuous (anti-)dual-BRST [(anti-)co-BRST] symmetry 
transformations from our earlier work [6, 10].} ($ s_{(a)d}\,(\dot{\lambda} - p_r) 
= 0,\,s_{(a)b}\,B  = 0$) and the nilpotency and absolute anticommutativity of  
$ s_{(a)d} $ are valid {\it off-shell} where we do not use any EL-EOM. The 
generators of the symmetry transformations (7) are\footnote{It will be noted 
that the Noether theorem yields the charges as $Q_d = B\bar{C} - (r - a)\, 
\dot{\bar{C}}$ and $Q_{ad} = BC - (r - a)\,\dot{C}$. These are re-expressed 
as (8) by using the EL-EOM (6).}
\begin{eqnarray}
Q_d = \dot{R}\, \bar{C} - R\, \dot{\bar{C}}  
\equiv B\,\bar{C} + \dot{B}\, \dot{\bar{C}}, \qquad
Q_{ad} =&\dot{R}\, C - R\, \dot{C}  
\equiv  B\,C + \dot{B}\, \dot{C}.
\end{eqnarray}
We note that these charges are nilpotent (i.e. $Q_{(a)d}^2 = 0 $) of 
order two and they are absolutely anticommuting ($Q_d\,Q_{ad} 
+ Q_{ad}\,Q_{d} = 0 $) in nature, namely;
\begin{eqnarray}
s_d\, Q_d &=& -\,i\,\lbrace Q_d,\, Q_d\rbrace = 0, \qquad\quad\quad\;\; 
s_d\, Q_{ad} = \,-i\, \lbrace Q_{ad},\, Q_d \rbrace = 0, \nonumber\\
s_{ad}\, Q_{ad} &=&  -\,i\,\lbrace Q_{ad}, Q_{ad}\rbrace = 0, 
\qquad\quad\quad s_{ad}\,Q_d = \,-\,i\, \lbrace Q_d, Q_{ad}\rbrace = 0,
\end{eqnarray}
when we use the equations of motion (6). We stress that the physicality 
criteria with the nilpotent and conserved (anti-)co-BRST charges 
$ Q_{(a)d}\mid phys> = 0 $ lead to the  annihilation of the physical 
states by the operator form of the first-class constraints of the 
theory (as was the case with such kind of criteria with the conserved 
and nilpotent (anti-)BRST charges).

The anticommutator $ (\lbrace s_b,\,s_{d}\rbrace = -\,\lbrace s_{ab},\,
s_{ad} \rbrace = s_w) $ of the (anti-)BRST and (anti-)co-BRST symmetry 
transformations leads to the definition of a unique\footnote{The transformations 
$s_w = \{s_b,\,s_d\}$ and ${\bar s}_w =\{s_{ad},\,s_{ab}\}$ look different in the 
beginning but it can be checked that $s_w + {\bar s}_w = 0$ when we use the
appropriate EL-EOM of our present theory.} bosonic symmetry $(s_w)$ in our theory 
[6, 10]. The transformations of variables under this symmetry are
\begin{eqnarray}
&& s_w\, p_r = i\,[\dot B - (r - a)] \equiv  i\,(\dot{B} - R),
\qquad  s_w\, ( r, \theta, C, \bar{C}, B) = 0,  \nonumber\\
&& s_w\,\lambda = i\,\Bigl [B + \frac{d}{dt}\,(r - a) \Bigr ]   
\equiv i\, (B + \dot{R}),\nonumber\\
&&s_w\, L_B = i\,\frac{d}{dt} \Bigl [B\,\frac{d}{dt}(r - a) - 
(r - a)^2 \Bigr ] \equiv  i\,\frac{d}{dt}( B\,\dot{R} - R^2 ), 
\end{eqnarray}
which demonstrate that the action integral $S = \int {dt\, L_B}$ 
remains invariant under the bosonic transformations ($s_w$). 
The conserved charge, corresponding to the above continuous 
symmetry transformations, is as follows:
\begin{eqnarray}
Q_w = i\, ( R^2 + B^2 ) \equiv i\,[B\,\dot R - R\,\dot B].
\end{eqnarray}
The conservation law of this charge can be proven by using 
the  the EOM (6).

We observe that the Lagrangian $ L_B $ remains invariant 
under the following ghost-scale symmetry transformations for the 
variables of our theory, namely;
\begin{equation}
C \longrightarrow e^{+1\,\Lambda}\, C, \quad \bar{C} 
\longrightarrow e^{-1\,\Lambda}\, \bar{C}, 
\quad \Phi \longrightarrow e^{0\Lambda}\, \Phi, \quad 
(\Phi = r,\, \theta,\, p_r, \lambda,\, B),
\end{equation}
where $\Lambda$ is a global parameter and numerals in the 
exponential denote the ghost number of the variables. The infinitesimal 
version of the above transformations is:
\begin{eqnarray}
s_g \, C = +\, C, \quad\quad s_g\, \bar{C} = -\, \bar{C}, 
\quad\quad s_g\, \Phi = 0, 
\quad\quad (\Phi = r,\,\theta,\, p_r,\, \lambda, \, B), 
\end{eqnarray}
where we have set, for the sake of brevity, the scale parameter 
(present in (12)) equal to one (i.e. $ \Lambda = 1$). The 
conserved charge corresponding to (13) is:
\begin{eqnarray}
Q_{g} = i\,(\bar{C}\,\dot{C} - \dot{\bar{C}}\,C), 
\qquad\qquad\qquad\qquad \dot{Q}_g = 0.
\end{eqnarray} 
The above charge is also the generator of transformations (13) as  
\begin{equation}
s_g\,C = +\,i \,\bigl[ C,\, Q_g \bigr] = +\, C, \qquad\qquad\qquad 
s_g\, \bar{C} = + \, i\, \bigl[ \bar{C},\,Q_g \bigr] = -\, \bar{C}.
\end{equation}
Similarly, the trivial ghost-scale transformations on the variables 
$\phi = r, \theta, B, \lambda, p_r $ can be written as
$s_g\,\phi = -\,i\, [\phi, \, Q_g] = 0$ because the variables 
$r, \lambda, p_r, \theta, B$ commute with the ghost variables of the 
charge $Q_g$. Thus, ultimately, we conclude that there are {\it six} 
continuous symmetries in the toy model (i.e. 1D rigid rotor) of our 
present example of Hodge theory [6].

\noindent
\section{Canonical Quantization: Normal Mode Expansions}

We note that the second term (i.e. $r^2\,\dot{\theta}^2 /2$) in 
the Lagrangian (2) does not contribute anything as far as the 
symmetries of the theory are concerned. For a definite kinetic 
energy of the rigid rotor, this term becomes a constant and, 
therefore, it can be ignored. In particular, if the angular 
velocity (i.e. $\dot{\theta}$) is constant, the term 
($r^2\,\dot{\theta}^2 /2$) becomes a constant (which could 
be a constant number). In view of these arguments, we ignore 
the second term of the Lagrangian. As pointed out earlier, 
the constraint-line of our theory is defined by the relations 
$(r - a)\approx 0$ and $d/dt\,(r - a)\approx 0 $ which are the 
first-class constraints on our theory. If we confine our system 
to evolve on this constraint-line, the equations of motion (6) 
would reduce to the following {\it simple} and nice-looking
form\footnote{It should be noted that the EOM (6) yield the relationship 
$\frac{d^2}{dt^2}\,(\dot{\lambda} - p_r) + (\dot{\lambda} - p_r) 
= 0$ without any approximation. These equations can be 
re-expressed as $\dddot{\lambda} + \dot{\lambda} - (\ddot{p_r} 
+ p_r) = 0$. One of its solutions of our interest is: $\ddot{\lambda}
+ \lambda = 0$ together with $ \ddot{p_r} + p_r = 0$ (see, also Appendix B). 
These relations are also derived as EL-EOM when we ignore the second term 
[$ (r^2\, \dot{\theta}^2 )/2 $] from the Lagrangian (2) of our theory 
(cf. Sec. 2).}:
\begin{eqnarray}
&& \ddot{C} + C = 0, \;\qquad\qquad\qquad \ddot{\bar{C}} + \bar{C} = 0, 
\qquad\qquad\qquad \ddot{\lambda} + \lambda = 0, \nonumber\\
&& \ddot{p_r} + p_r = 0, \qquad\qquad\qquad \ddot{R} + R = 0, 
\qquad\qquad\qquad \ddot B + B = 0.
\end{eqnarray}
We re-emphasize that the above EL equations of motion are valid 
for a rigid rotor with a constant kinetic energy moving on a 
circle of radius $r = a$ at {\it all times} during its physical 
evolution which is described by the following Lagrangian 
\begin{equation}
L_B \longrightarrow  L^{(0)}_B = \dot r\,p_r - \lambda \,(r - a) + B\,
(\dot{\lambda} - p_r) + \frac{1}{2}\,B^2 - i\, \dot{\bar C}\,\dot C 
+ i\, \bar C\, C.
\end{equation}
This is the Lagrangian we shall focus on for the rest of our 
discussions.

The above EL equations of motion (16) have their solutions in terms 
of the mode expansions (see e.g. [9]) where the creation and 
annihilation operators appear at the quantum level. These mode 
expansions, in their explicit forms, are as follows
\begin{eqnarray}
&& R(t) = \frac{1}{\sqrt{2}}\, \left[ s\, e^{-it} 
+ s^{\dagger}\,e^{+it} \right], 
\quad \lambda(t) = \frac{1}{\sqrt{2}}\, \left[ d\, e^{-it} 
+ d^{\dagger}\,e^{+it} \right],\nonumber\\
&& C(t) = \frac{1}{\sqrt{2}}\, \left[ c\, e^{-it} 
+ c^{\dagger}\,e^{+it} \right], \quad \bar{C}(t) = \frac{1}{\sqrt{2}}
\, \left[ \bar{c}\, e^{-it} + \bar{c}^{\dagger}\,e^{+it} \right], \nonumber\\
&& p_r(t) =  \frac{1}{\sqrt{2}}\, \left[ k \, e^{-it} 
+ k^{\dagger}\,e^{+it} \right], \;\; B(t) = \frac{1}{\sqrt{2}}\, 
\left[ l\, e^{-it} + l^{\dagger}\,e^{+it} \right],
\end{eqnarray}
where the time-independent dagger and non-dagger operators are the creation 
and annihilations operators. It is clear,  from the Lagrangian (17), 
that we have the following canonically conjugate momenta in our present 
theory, namely;
\begin{eqnarray}
\Pi_{(C)} = +\,i\,\dot{\bar{C}}, \qquad\quad \Pi_{\bar{(C)}} = 
- i\,\dot{C}, \qquad\quad \Pi_{(\lambda)} = B, 
\qquad\qquad \Pi_{(R)} = p_r,  
\end{eqnarray}
which lead to the basic canonical brackets as
\begin{eqnarray}
[R, \Pi_{(R)}] = i, \quad\quad [\lambda, \,B] = i, 
\quad\quad \lbrace C, \, \Pi_{(C)} \rbrace = i, 
\quad\quad \lbrace \bar{C}, \, \Pi_{\bar{(C)}} \rbrace = i, 
\end{eqnarray}
and the rest of the brackets are zero. It is to be noted that 
the above (anti)commutators reduce to the following forms in 
terms of the explicit variables, namely;
\begin{eqnarray}
\bigl[R (t),\,p_r (t) \bigr] = i, \,\,\, [\lambda (t),\, B (t) ] =  i, 
\,\,\, \lbrace C(t),\, \dot{\bar{C}} (t) \rbrace = 1, 
\,\,\, \lbrace \bar{C} (t), \, \dot{C} (t) \rbrace = -\,1.
\end{eqnarray}
We shall concentrate on (21) for the rest of our central analysis and arguments. 
The above (anti)commutators (21) can be re-expressed in terms of 
the creation and annihilation operators of the mode expansions (18) 
as 
\begin{eqnarray}
\bigl[ s,\, k^\dagger \bigr] = i \equiv \bigl[ s^\dagger,\, k \bigr], 
\,\,\,\lbrace c,\, \bar{c}^\dagger \rbrace =  -\,i,\,\,\, \lbrace \bar{c},
\, c^\dagger \rbrace = +\,i, \,\,\, \bigl[d,\, l^\dagger \bigr] = +\,i \equiv 
\bigl[d^\dagger, \,l\bigr],
\end{eqnarray}
and the rest of the (anti)commutators are zero. In other words, 
we have primarily {\it four} non-vanishing (anti)commutators at 
the {\it quantum} level and rest of {\it all} the (anti)commutators 
are zero (see, Appendix A below) as far as the canonical quantization scheme 
is concerned.

We would like to lay emphasis on the fact that we have utilized the 
spin-statistics theorem and the mathematical definition of the canonical 
conjugate momenta to derive the basic canonical (anti)commutators which 
quantize our system of a one (0 + 1)-dimensional rigid rotor. There has 
{\it not} been any urgent need to exploit the idea of normal ordering as 
we have not expressed the Hamiltonian of our present theory in terms of 
the creation and annihilation operators. However, the latter idea is also 
one of the important ingredients of the standard canonical quantization scheme 
for a given physical system. We shall see that, in our forthcoming sections,
this idea of normal ordering would play an important role in the context of the 
\textit{proper} physical expressions for the Noether conserved charges of our theory.

\noindent
\section{Ghost Symmetries: Basic Canonical Brackets}

Using the mode expansions (18), we can express the conserved charge $Q_g$  
in terms of the creation and annihilation operators as
\begin{eqnarray}
Q_g = \bar{c}^\dagger\,c - \bar{c}\,c^\dagger \;\;\Longrightarrow \quad 
:Q_g:\; = \, \bar{c}^\dagger\,c + c^\dagger\,\bar{c},
\end{eqnarray}
where we have used the idea of normal ordering to re-arrange all the 
creation operators to the left and annihilation operators to 
the right so that the above conserved charge $Q_g$ could make some 
physical sense for our present theory.

We exploit now the virtues of (15) in deriving the anticommutators 
amongst the creation and annihilation operators of the expansion for 
$C(t)$ and $\bar{C}(t)$. Plugging in the expansion for $C(t)$ in (15), 
we obtain the following
\begin{eqnarray}
&& \lbrace c,\,\bar{c} \rbrace \;\;\;=  \; \lbrace c,\, c^\dagger \rbrace \; 
=  \lbrace c, \, c\rbrace = 0, 
\qquad \lbrace c, \bar{c}^\dagger\rbrace = -\,i, \nonumber\\
&& \lbrace c^\dagger,\, \bar{c}^\dagger \rbrace \,= \; \lbrace c^\dagger,
\, c \rbrace \;= \lbrace c^\dagger,\, c^\dagger \rbrace = 0, \quad
\lbrace c^\dagger,\,\bar{c}\rbrace = +\,i.
\end{eqnarray}
Similarly, the substitution of expansion for $\bar{C}(t)$, leads to 
\begin{eqnarray}
&& \lbrace \bar{c},\, c^\dagger \rbrace \;\;=\;  \lbrace\bar{c},\, 
c \rbrace \quad = \; \lbrace \bar{c},\, \bar{c} \rbrace = 0, 
\qquad\;\; \lbrace \bar{c},\, c^\dagger \rbrace = + \,i,\nonumber\\
&& \lbrace \bar{c}^\dagger, \,\bar{c} \rbrace \;\;= 
\;\lbrace\bar{c}^\dagger, \,c^\dagger \rbrace \;=\; 
\lbrace \bar{c}^\dagger,\,\bar{c}^\dagger \rbrace = 0, 
\quad\;\; \lbrace \bar{c}^\dagger, \, c \rbrace = - i,
\end{eqnarray}
where we have compared the coefficients of the exponentials\footnote{This 
is due to the fact that the exponentials $e^{-it}$ and $e^{+it}$ are linearly 
independent of each-other as they are the solutions of the generic EOM for 
the variable $\Psi$: $(\frac {d^2}{{dt}^2} + 1)\,\Psi = 0$ where $\Psi = C, 
\bar C.$ The linear independence can be proven by showing that the 
Wronskian (for the above second-order differential equation) turns out 
to be non-zero for these solutions.} $e^{-it}$ and $e^{+it}$ from the 
l.h.s. and r.h.s. of (15). The bottom-line of this discussion is the observation 
that the non-vanishing brackets from (15) are $\lbrace c,\, \bar{c}^\dagger \rbrace 
= -\, i$ and  $\lbrace \bar{c},\, c^\dagger \rbrace= +\, i$ which 
are exactly same as the ones derived from the usual canonical method 
of quantization (cf. Sec. 3 for details).

We now concentrate on the trivial ghost-scale transformations
\begin{equation}
s_g\, \Phi = i\, \bigl[\Phi,\, Q_g \bigr] = 0, 
\qquad\qquad\qquad\qquad \Phi =B,\, R, \,\lambda,\, p_r.
\end{equation}
Using the expansions for $Q_g$ (from (23)) and the mode 
expansions for $\lambda, \,R,\, p_r, B$ from (18), it is 
evident that the relation (26) leads to the derivation of the following:
\begin{eqnarray}
&& [l, c] = 0, \qquad\; [l, c^\dagger] = 0, \qquad\;\;\, [l, \bar{c}] = 0, 
\qquad\;\, [l, \bar{c}^\dagger] = 0, \nonumber\\
&& [l^\dagger, c] = 0, \qquad [l^\dagger, c^\dagger] = 0, \qquad\, 
[l^\dagger, \bar{c}] = 0, \qquad [l^\dagger, \bar{c}^\dagger] 
= 0, \nonumber\\
&& [s, c] = 0, \qquad\; [s, c^\dagger] = 0, \qquad\;\; [s, \bar{c}] = 0, 
\qquad\; [s, \bar{c}^\dagger] = 0, \nonumber\\
&& [s^\dagger, c] = 0, \qquad [s^\dagger, c^\dagger] = 0, \qquad 
[s^\dagger, \bar{c}] = 0, \qquad [s^\dagger, \bar{c}^\dagger] 
= 0, \nonumber\\
&& [d, c] = 0, \qquad\; [d, c^\dagger] = 0, \qquad\;\, [d, \bar{c}] = 0, 
\qquad\; [d, \bar{c}^\dagger] = 0, \nonumber\\
&& [d^\dagger, c] = 0, \qquad [d^\dagger, c^\dagger] = 0, \qquad 
[d^\dagger, \bar{c}] = 0, \quad\;\;\; [d^\dagger, \bar{c}^\dagger] 
= 0,  \nonumber\\
&& [k, c] = 0, \qquad\; [k, c^\dagger] = 0, \qquad\,\; [k, \bar{c}] 
= 0, \qquad\; [k, \bar{c}^\dagger] = 0, \nonumber\\
&& [k^\dagger, c] = 0, \qquad [k^\dagger, c^\dagger] = 0, \qquad 
[k^\dagger, \bar{c}] = 0, \qquad [k^\dagger, \bar{c}^\dagger] 
= 0.
\end{eqnarray}
Ultimately, we conclude that, we have obtained {\it all} the brackets 
that emerge from the ghost-scale transformations (13) and the non-vanishing 
brackets are the anticommutators $\lbrace c,\, \bar{c}^\dagger \rbrace = -\,i$ 
and $\lbrace \bar{c},\, c^\dagger \rbrace = +\,i$ which are consistent 
with the canonical anticommutators derived in Sec. 3. We lay stress on the 
fact that we have {\it not} used the definition of the canonical conjugate 
momenta w.r.t. $C$ and $\bar{C}$ in our derivations of the non-vanishing 
canonical anticommutators $\lbrace c,\, \bar{c}^\dagger \rbrace = -\,i $ 
and $\lbrace \bar{c},\, c^\dagger \rbrace = +\,i$. Instead, we have 
exploited the idea of symmetry principles where the continuous symmetries and 
their generators play the decisive roles. We observe that the ghost-scale 
symmetry {\it alone} does not produce the non-vanishing brackets 
$ [ s, k^\dagger] = i \equiv [s^\dagger, k]$ and  $ [ d, l^\dagger] 
= i \equiv [d^\dagger, l]$.
Thus, other continuous symmetries of the theory are required for the complete 
derivation of {\it all} the canonical basic brackets.

\noindent
\section{Nilpotent (Anti-)BRST Symmetries: Fundamental (Anti)commutators}

From the expressions for the (anti-)BRST charges $Q_{(a)b}$, it is 
clear that these can be expressed in terms of the mode expansion (cf. (18)) as
\begin{eqnarray}
:Q_b: = (s^\dagger\, c + c^\dagger \,s) 
\equiv i\,(c^\dagger\,l - l^\dagger\, c), \quad
:Q_{ab}: = (s^\dagger\, \bar{c} + \bar{c}^\dagger \, s) 
\equiv i\,(\bar{c}^\dagger\, l - l^\dagger\, \bar{c}),
\end{eqnarray}
where we have used the equivalent expressions for (anti-)BRST charges 
as
\begin{eqnarray}
Q_b = B\, \dot{C} - \dot{B}\,C \equiv \dot{R}\, \dot{C} + R\,C, \qquad\,\,
Q_{ab} = B\, \dot{\bar{C}} - \dot{B}\, \bar{C} \equiv \dot{R}\,\dot{\bar C} + R\,\bar{C},
\end{eqnarray}
and taken the normal ordering into consideration in (28).
The conservation law on $Q_{(a)b}$ compels that these charges should 
be independent of time. In other words, we note that $\dot{Q}_{(a)b} = 0$ 
turns out to be true if we use
$ \ddot{R} + R = 0,\, \ddot{C} + C = 0,\,   \ddot{\bar{C}} + \bar{C} = 0, 
\ddot{B} + B = 0$.
The above forms of the normal ordered charges (28) are automatically conserved as the 
terms present in the above expressions are time-independent by their very
definitions. We would like to emphasize that the Noether conserved charges emerge from 
the action principle where the mathematical definition 
of the canonical conjugate momenta does {\it not} play any role. Thus, in our discussions, we have
not used the definition of canonical conjugate momentum.

We observe that $s_{(a)b}\, R = 0$ (since $s_{(a)b}\, r = 0$ in (3)). 
Thus, it is clear that $s_{(a)b}\, R = -\,i\,[R, \, Q_{(a)b}] = 0$. 
Taking the mode expansion for $R(t)$ from (18) and that for the $Q_{(a)b}$ 
from (28), we find the creation and annihilation operators $s$ and $s^\dagger$ 
commute with all the creation and annihilation operators present in (28). 
In other words, we have the following: 
\begin{eqnarray}
&& [s,\,s^\dagger] = [s,\,c] = [s,\,c^\dagger] = [s^\dagger, c] 
= [s^\dagger,\, c^\dagger] = 0,\nonumber\\
&& [s,\,l] = [s^\dagger,\,l] = [s,\,l^\dagger] 
= [s^\dagger,\,l^\dagger] = 0,\nonumber\\
&& [s,\,\bar{c}^\dagger] = [s^\dagger,\, \bar{c}]  
= [s^\dagger, \bar{c}^\dagger] = [s,\,\bar{c}] = 0. 
\end{eqnarray}
Thus, we have obtained a vanishing set of commutators from 
$s_{(a)b}\,R = 0 = -\,i\,[R,\, Q_{(a)b}]$. Now, we concentrate 
on the transformations $s_b\, C = 0$ and $s_{ab}\, \bar{C} = 0$. 
These, finally, imply the following in terms of the (anti-)BRST charges, 
namely;
\begin{equation}
s_b\,C = -\, i\, \lbrace C,\, Q_b \rbrace = 0, \qquad
s_{ab}\, \bar{C} = -\,i\, \lbrace \bar{C}, \, Q_{ab} \rbrace = 0.
\end{equation}
Using the mode expansions from (18) and exploiting the explicit 
expressions
for $Q_{(a)b}$ (from (28)), we obtain the following independent basic brackets:
\begin{eqnarray}
&& \lbrace c, \, c^\dagger \rbrace = [c, l] = [c, l^{\dagger}] 
=  \lbrace c, \, c \rbrace = 0,\nonumber\\
&& \lbrace {\bar c}, \, {\bar c}^\dagger \rbrace = [\bar c, l] 
= [\bar c, l^{\dagger}] =  \lbrace {\bar c}, \, {\bar c} \rbrace = 0,
\end{eqnarray}
where we have used $Q_b = B\, \dot{C} - \dot{B}\,C 
=  i\,(c^\dagger\,l - l^\dagger\, c)$ and 
$ Q_{ab} = B\, \dot{\bar{C}} - \dot{B}\, \bar{C} 
= i\,(\bar{c}^\dagger\, l - l^\dagger\, \bar{c})$
because these are the forms that can be used for 
the computation of $s_b\, \bar C = i\, B,$ $s_{ab}\, C = -\,i\, B $.
Thus, once again, we have obtained some vanishing (anti)commutators 
from the transformations $s_b\, C = 0$ and $s_{ab}\,\bar{C} = 0$ 
by exploiting the idea of symmetry generators. 

Now, we set out to obtain the (non-)vanishing brackets from the relations
$s_b\, p_r = -\, C$ and $s_{ab}\, p_r = -\, \bar{C}$ (that are present 
in (3)), as:
\begin{equation}
s_b\, p_r = -\, i\, \big[p_r, Q_b\big] = -\, C,
\qquad s_{ab}\, p_r = -\, i\, \big[p_r, Q_{ab}\big] = -\, \bar{C}.
\end{equation}
Using the expansions from (18) and expressions (28), we obtain
\begin{eqnarray}
&& [s, k^\dagger] = i = [s^\dagger, k], \; [k,s] 
= [k^\dagger, s^\dagger] = 0, \nonumber\\
&& [k, c] \;= [k, \bar{c}]  \,\;= [k, c^\dagger] 
\;= [k, \bar{c}^\dagger] = 0, \nonumber\\
&& [k^\dagger, c] = [k^\dagger, \bar{c}] 
= [k^\dagger, \bar{c}^\dagger] = [k^\dagger, c^\dagger] = 0,
\end{eqnarray}
which shows that the non-vanishing (and consistent with the canonical 
brackets (22)) are the brackets $[s, k^\dagger] = i$ and its 
Hermitian conjugate $[s^\dagger, k] = i$. The rest of the brackets 
are zero because the momentum operator $p_r$ commutes with (anti-)ghost 
operators. Similar exercise with the symmetry transformations
\begin{equation}
s_b\, \lambda = -\,i\,\big[\lambda, Q_b \big] = \dot{C},
\qquad s_{ab}\, \lambda = -\,i\,\big[\lambda, Q_{ab} \big] = \dot{\bar{C}},
\end{equation}
leads to the following basic (anti)commutators at the level of 
creation and annihilation operators:
\begin{eqnarray}
&& [d,l^\dagger] = i = [d^\dagger, l], \; [d,l] = 0 
= [d^\dagger, l^\dagger], \nonumber\\
&& [d,c] \;\,= \;[d,c^\dagger] \;=\, [d,\bar{c}] 
\;\,= [d, \bar{c}^\dagger] = 0, \nonumber\\
&& [d^\dagger,c] = \;[d^\dagger, c^\dagger] =\, [d^\dagger, \bar{c}] 
= [d^\dagger, \bar{c}^\dagger] = 0.
\end{eqnarray}
We note that the non-vanishing bracket $[d, l^\dagger] = i$
and its Hermitian conjugate $[d^\dagger, l] = i$ are
same as the canonical brackets listed in (22). We focus
on the transformations
\begin{eqnarray}
s_b\, \bar{C} = -\,i\,\lbrace \bar{C}, Q_b \rbrace = i\, B, \qquad
s_{ab}\, C = -\, i\, \lbrace C, Q_{ab} \rbrace = -\, i\, B,
\end{eqnarray}
and perform the earlier exercise to obtain the non-vanishing 
anticommutators $ \lbrace \bar{c}, c^\dagger \rbrace = i,$ $  
\lbrace c, \bar{c}^\dagger \rbrace = -\,i $ 
that are consistent with the canonical brackets (22).
The vanishing brackets from our present exercise are as follows:
\begin{eqnarray}
&& [c,l] \;\,= \;[c, l^\dagger] \quad= \;[c^\dagger, l] \quad 
= \; [c^\dagger, l^\dagger] 
= 0, \nonumber\\
&& [\bar{c}, l] \,\;=\; [\bar{c}, l^\dagger] \quad 
=\;  [\bar{c}^\dagger, l] 
\quad = \; [\bar{c}^\dagger, l^\dagger] = 0, \nonumber\\
&& \lbrace c, c \rbrace = \lbrace c^\dagger, c^\dagger \rbrace 
\;= \lbrace \bar{c}, \bar{c} \rbrace 
\quad\, = \; \lbrace \bar{c}^\dagger, \bar{c}^\dagger \rbrace  = 0.  
\end{eqnarray} 
We emphasize that the above brackets are also consistent with the canonical 
brackets (22). As pointed out earlier, we have to use here the forms
of the conserved and nilpotent BRST and anti-BRST charges as:
$Q_b = B\, \dot{C} - \dot{B}\,C 
=  i\,(c^\dagger\,l - l^\dagger\, c)$ and 
$ Q_{ab} = B\, \dot{\bar{C}} - \dot{B}\, \bar{C} 
= i\,(\bar{c}^\dagger\, l - l^\dagger\, \bar{c})$.
We concentrate on the trivial transformations $s_b\,B  = 0$
and $s_{ab}\, B = 0$. These lead to the 
derivation of the following vanishing brackets (with both 
the expressions for $Q_b$ and $Q_{ab}$ listed in (28)), namely;
\begin{eqnarray}
&& [l, c] = [l, c^\dagger] = [l^\dagger, c] 
= [l^\dagger, c^\dagger] =  [l^\dagger, s] 
= [l^\dagger, s^\dagger]  = [l, l^\dagger] = 0, \nonumber\\
&& [l, \bar{c}] = [l, \bar{c}^\dagger] = [l^\dagger, \bar{c}] 
= [l^\dagger, \bar{c}^\dagger] = [l, s] = [l, s^\dagger] = 0. 
\end{eqnarray}
We, finally, conclude that all the vanishing as well as non-vanishing 
canonical quantum  brackets (i.e. basic (anti-)commutators) of the 
standard canonical quantization scheme can be derived from the virtues 
of symmetry principles {\it alone} where the mathematical 
definition of the canonical conjugate momenta w.r.t. all the dynamical 
variables is {\it not} required.

\noindent
\section{(Anti-)co-BRST Symmetries: Basic Brackets}

Using the expansions of (18), we note that the (anti-)co-BRST
charges $ Q_{(a)d} $ (i.e. $ Q_d = B\, \bar{C} + \dot{B}\, 
\dot{\bar{C}} \equiv \dot{R} \, \bar{C}  - R\, \dot{\bar{C}}$
and $ Q_{ad} = B\, C + \dot{B}\, \dot{C}
\equiv \dot{R}\, C - R\, \dot{C} $) can be expressed as:
\begin{eqnarray}
:Q_d: \, = \, l^\dagger\, \bar{c} + \bar{c}^\dagger\, l
\equiv i\, (s^\dagger\, \bar{c} - \bar{c}^\dagger\,s),\qquad
:Q_{ad}: \, = \, l^\dagger\, c + c^\dagger\, l
\equiv i\, (s^\dagger\, c - c^\dagger\,s), 
\end{eqnarray}
where the process of normal ordering has been adopted. We are 
in a position now to proceed in the manner that has been followed 
in our previous section. It is trivial to note that
$s_{(a)d}\,(R, B) = 0,\, s_d\, \bar{C} = 0, \, 
s_{ad}\,C = 0$. These can be expressed in terms of $Q_{(a)d}$ as 
\begin{eqnarray}
s_{(a)d}\, B &=& -\, i\, \big[B, Q_{(a)d}\big] = 0, \qquad
s_{(a)d}\, R = -\, i\, \big[R, Q_{(a)d}\big] = 0, \nonumber\\
s_d \, \bar{C} &=& -\,i\, \lbrace \bar{C}, Q_d \rbrace = 0,\qquad
\quad\;\, s_{ad}\, C = -\,i\, \lbrace C, Q_{ad} \rbrace = 0.
\end{eqnarray}  
The above brackets lead to the following basic (anti)commutators
amongst the creation and annihilation operators of the normal mode
expansions (18), namely;
\begin{eqnarray}
&& [s, l^\dagger] = [s, l] = [s, \bar{c}^\dagger] = [s,\bar{c}] 
=  [s^\dagger, l^\dagger] = [s^\dagger, l] 
= [s^\dagger, \bar{c}^\dagger] = [s^\dagger,\bar{c}] = 0, \nonumber\\
&& [s, s^\dagger ]  = \lbrace c, \, c \rbrace = \lbrace c, c^\dagger \rbrace 
=  \lbrace c^\dagger, c^\dagger \rbrace = \lbrace \bar{c}, \, \bar{c} \rbrace = 
\lbrace \bar{c}, \bar{c}^\dagger\rbrace = \lbrace \bar{c}^\dagger, 
\bar{c}^\dagger\rbrace = 0,
\end{eqnarray}
where we have quoted {\it only} the independent canonical quantum brackets
that emerge from 
$s_{(a)d}\,\phi = -\, i\, \big[\phi, Q_{(a)d}\bigr]_\pm = 0 $ 
where ($\pm$) signs on the square bracket correspond to 
the (anti)commutator for the generic variables
$\phi = R,\, B,\, C,\,\bar{C}$ being (fermionic) bosonic in nature for our theory 
under consideration.

We next focus on the derivation of basic brackets from the symmetry 
transformations $s_d \, \lambda  = -\,i\, \big[\lambda, Q_d\big] = \bar{C}$
and $s_{ad}\, \lambda  = -\,i\, \big[\lambda, Q_{ad} \big] = C $ where
the conserved charges $Q_d = B\,\bar{C} + \dot{B}\,\dot{\bar{C}}$
and $Q_{ad} = B\,C + \dot{B}\, \dot{C}$  play important roles. 
Using the expansions from (18) and appropriate expressions for $Q_{(a)d}$
from (40), we obtain the following (non-)vanishing basic 
(anti)commutators amongst the creation and annihilation operators, namely;
\begin{eqnarray}
&& [d, l^\dagger] \;= i = \;[d^\dagger, l], \quad
[d, c] \;\;= [d, \bar{c}] \;= [d, c^\dagger] 
\;= [d, \bar{c}^\dagger] = 0,  \nonumber\\
&& [d, l] \;\; = 0 = \; [d^\dagger, l^\dagger],\;\;
[d^\dagger, c] \,= [d^\dagger, \bar{c}] 
= [d^\dagger, c^\dagger] = [d^\dagger, \bar{c}^\dagger]
= 0.
\end{eqnarray}
Thus, we note that the non-vanishing basic brackets $[d, l^\dagger] 
= i $ and its Hermitian conjugate $[d^\dagger, l] = i$ are
consistent with the canonical brackets defined in our Sec. 3. Similar 
exercise for the transformations $s_d\, C = i\, R \equiv i\, (r - a)$
and $s_{ad}\, \bar{C} = -i\, R = -\, i\, (r - a)$ with the (anti-)co-BRST 
charges, written in the following manner, namely;
\begin{eqnarray}
s_d\, C &=& -\, i\, \lbrace C, Q_d \rbrace 
\;\;\equiv -\, i\,\lbrace C, \dot{R}\, \bar{C} -\, R\,\dot{\bar{C}} \rbrace 
= i\, R, \nonumber\\
s_{ad}\bar{C} &=& -\, i\, \lbrace C, Q_{ad} \rbrace 
\,\equiv -\, i\,\lbrace \bar{C}, \dot{R}\, C - \,R\,\dot{C} \rbrace 
= -\, i\, R,
\end{eqnarray}
leads to the derivation of the following basic (non-)vanishing brackets:
\begin{eqnarray}
&& \lbrace c, \bar{c}^\dagger \rbrace \;= - \, i, \,\quad
\lbrace c^\dagger, \bar{c} \rbrace = + \, i,\quad\;\;\, \lbrace \bar{c}^\dagger, 
c^\dagger \rbrace = 
\lbrace c^\dagger, \bar{c}^\dagger \rbrace = 0, \nonumber\\
&& [c, s] \quad = \;[c, s^\dagger ] \quad  
= \; \lbrace c, \bar{c} \rbrace 
= 0, \quad  [c^\dagger, s] \;\; = \;\, [c^\dagger, s^\dagger] = 0, \nonumber\\
&& [\bar{c}, s] \quad =\; [\bar{c}, s^\dagger ] \quad  
= \; \lbrace \bar{c}, \bar{c} \rbrace = 0, \quad
[\bar{c}^\dagger, s] \;\; = \; [\bar{c}^\dagger, s^\dagger ] \, 
= 0.
\end{eqnarray}
Thus, we observe that the symmetry transformations $s_d\, C 
= i\, R$ 
and $s_{ad}\, \bar{C}  = -\, i\, R$ produce the non-vanishing 
anticommutators between the creation and annihilation operators for 
the (anti-)ghost variables as:
$\lbrace c, \bar{c}^\dagger \rbrace = -\, i$ and 
$\lbrace \bar{c}, c^\dagger \rbrace = +\, i$ which are consistent 
with such basic anticommutators defined in the case of canonical 
method of  quantization (cf. Sec. 3). Finally, we concentrate on 
the transformations $s_d\, p_r = \dot{\bar{C}}$ and 
$s_{ad}\, p_r = \dot{C}$. These can be written (in terms of the (anti-)co-BRST 
charges $Q_{(a)d}$) as:
\begin{eqnarray}
s_d\, p_r &=& -\,i\,[p_r, Q_d] \;\;\equiv \;-\,i\,\big[p_r, \, 
\dot{R}\,\bar{C} - R\, \dot{\bar{C}}\big] = \dot{\bar C}, \nonumber\\
s_{ad}\, p_r &=& -\,i\,[p_r, Q_{ad}] \;\equiv \; -\,i\,
\big[p_r, \, \dot{R}\,C - R\,\dot{C}\big] = \dot{C}.
\end{eqnarray}
Plugging in the expansions from (18) and appropriate forms
(i.e. $Q_d = i\,(s^\dagger\, \bar{c} - \bar{c}^\dagger\,s),$ $
 Q_{ad} = i\,(s^\dagger\, c - c^\dagger\,s)$) of the conserved (anti-)co-BRST 
 charges $Q_{(a)d}$,
we obtain the following fundamental (anti)commutators amongst the 
creation and annihilation operators:
\begin{eqnarray}
&&[s,\,k^\dagger ] = i = [s^\dagger,\, k], \quad [k,\,\bar c] 
= [k,\,{\bar c}^\dagger] = [k,\, s] = [k^\dagger \, , \bar c] = 0,\nonumber\\
&& [k,\, c] = [k,\,c^\dagger] = [k^\dagger , \, c^\dagger] = [k^\dagger ,\, c] 
= [k^\dagger \, , {\bar c}^\dagger] = [k^\dagger\, , s^\dagger] =  0.
\end{eqnarray}
These (non-)vanishing (anti)commutators establish that the non-vanishing 
canonical brackets are $[s,\,k^\dagger ] = i$ and $[s^\dagger,\, k] = i$. 
These are consistent with such canonical brackets derived in Sec. 3. 
Thus, we conclude that all the basic brackets, derived from the (anti-)co-BRST 
charges and their corresponding symmetries, are consistent with the canonical 
brackets (i.e. (anti-)commutators) defined in Sec. 3. by the standard canonical method.

\noindent
\section{Bosonic Symmetries: Fundamental Brackets}

We devote time on the derivation of the basic canonical brackets that 
emerge from the symmetry transformations generated
by the bosonic conserved charge $Q_w = i\,(R^2 + B^2)$ (cf. Eq. (11)) 
which can be re-expressed, using the equations of motion (6), as
\begin{eqnarray} 
Q_w = i\,[B\,\dot R - \dot B\,R] \equiv i\,(R^2 + {\dot R}^2) 
\equiv i\,(B^2 + {\dot B}^2).
\end{eqnarray}
The above expansions can be written, in terms of the mode expansion (18), as follows:
\begin{eqnarray}
&&Q_w = (l^\dagger \, s - l\, s^\dagger) \quad \,\,\, \,\Longrightarrow
\,\,\,\,\, :Q_w:\,\,\, =\, (l^\dagger \, s - s^\dagger \,l),\nonumber\\
&& Q_w = i\,(s^\dagger \, s + s\, s^\dagger)\,\,\,\,\, \,\Longrightarrow
\,\,\,\,\, :Q_w:\,\,\, = \,2\,i\,s^\dagger\, s,\nonumber\\
&& Q_w = i\,(l^\dagger \, l + l\, l^\dagger)\quad\;\; \Longrightarrow
\,\,\,\,\, :Q_w:\,\,\, = \,  2\,i\,l^\dagger\, l,
\end{eqnarray}
where the procedure of normal ordering has been adopted in the last 
forms of $Q_w$. These expressions would be suitably used for our computations  
of the basic canonical brackets from the symmetry principles
where the \textit{appropriate} normal ordered expression for
$Q_w$ would be utilized as the generator for the bosonic symmetry 
transformations.

We note, from the bosonic symmetry transformations (10), that {\it only} 
the transformations $s_w\, p_r$ and $s_w\, \lambda$ exist
and rest of the variables of the theory do {\it not} transform at all. 
In  particular, we observe that, the (anti-)ghost variables do not transform 
under $s_w$. We would also like to state a few words on the forms of 
the non-vanishing transformations $s_w \,p_r$ and $s_w \,\lambda$ (cf. (10)) 
which can be re-expressed as:   
\begin{eqnarray}
s_w \,p_r &=& i\,(\dot B - R) \equiv - 2\,i\,R \,\equiv 2\,i\,\dot B,\nonumber\\
s_w \,\lambda &=& i\,(\dot R + B) \equiv \quad  2\,i\,B \equiv 2\, i\, \dot R,
\end{eqnarray}
by using EOM (6). It can be checked that, the following combinations: 
\begin{eqnarray}
&&s^{(1)}_w\,p_r = -\,2\,i\,R, \qquad \qquad s^{(1)}_w\,\lambda = 2\,i\,B,\nonumber\\
&& s^{(2)}_w\,p_r = 2\,i\,\dot B,\qquad \qquad\quad s^{(2)}_w\,\lambda = 2\,i\,\dot R,
\end{eqnarray}
are the symmetry transformations for the Lagrangian (17) and its corresponding action 
$S = \int \, dt\,L^{(0)}_B$ because we observe that the following is true, namely;
\begin{eqnarray}
s^{(1)}_w\,L^{(0)}_B = i\, \frac{d}{dt}\,\big(B^2 - R^2\big),\qquad s^{(2)}_w\,L^{(0)}_B 
= i\, \frac{d}{dt}\,\big(2\,\dot R\,B - R^2 - B^2\big).
\end{eqnarray}
Both the above bosonic symmetry transformations lead to the derivation of 
the conserved Noether charge as $Q_w = i\, \big(B^2 + R^2\big) $ which is also 
quoted in (11). The noteworthy point is that any other 
combinations of (50) are {\it not} found to be the symmetry of the Lagrangian
$L^{(0)}_B$ and the corresponding action (i.e. $S =\int dt\,L^{(0)}_B$).

Now we dwell a bit on the derivation of the canonical basic brackets from 
the symmetry transformations (51) and the conserved charge $Q_w$ defined in (49). 
These can be written as 
\begin{eqnarray}
s^{(1)}_w\,p_r &=& -\, i\, \big[p_r,\, Q_w \big] = -\, 2\,i\, R\nonumber\\ 
&\Longrightarrow&  -\,i\, \big[p_r, \, 2\,i\,s^\dagger\,s\big] = \frac{-2\,i}{\sqrt{2}}
\,\big(s\,e^{-i\,t} + s^\dagger\,e^{+i\,t}\big).
\end{eqnarray}
The comparison of the coefficients of $e^{-\,i\,t}$ and $e^{+\,i\,t}$ from the l.h.s. 
and r.h.s. 
leads to the following (non-)vanishing basic canonical brackets:
\begin{equation}
[k,\, s] = [k^\dagger,\,s^\dagger] = 0,\qquad \quad [k,\,s^\dagger] = i = [k^\dagger, \,s]. 
\end{equation}
It is to be noted that, even though the transformations $s^{(2)}_w$, are \textit{also} 
symmetry transformations for the action $S = \int\, dt\,L^{(0)}_B$, these 
transformations are {\it not}  interesting to us. Let us now concentrate on 
the following bosonic symmetry transformations:
\begin{equation}
s^{(1)}_w\, \lambda = -\,i\,\big[\lambda, \,Q_w\big] = 2\,i\,B \Longrightarrow 
-\,i\,\big[\lambda,\, 2\,i\,l^\dagger\,l \big] = \frac{2i}{\sqrt{2}}
\,\big(b\,e^{-\,i\,t} + l^\dagger\,e^{+\,i\,t} \big).
\end{equation}
Plugging in the expansion for $\lambda$ from (18) and taking the appropriate 
form of $Q_w = 2\,i\,l^\dagger\,l$ from (49), 
we obtain the following (non-)vanishing basic brackets:
\begin{equation}
[d,\,l] = [d^\dagger,\, l^\dagger] = 0, \,\,\quad\qquad [d,\,l^\dagger] = i 
= [d^\dagger, \,l].
\end{equation}
Thus, we point out that we have derived the non-vanishing brackets as 
$[d,\,l^\dagger] = i = [d^\dagger, l] $ which are in full agreement 
with the canonical brackets derived in Sec. 3, (cf. (22)). We re-emphasize 
that even though $s^{(2)}_w$ exists as a symmetry of the Lagrangian $L^{(0)}_B$ 
and corresponding action, it is not interesting for our purpose. 
Thus, we conclude that there is a {\it unique} bosonic symmetry $s^{(1)}_w\,p_r
= -\,2i\,R,$ $ s^{(1)}_w\,\lambda = 2i\,B, \, s^{(1)}_w\,\big(R,\,C,\, \bar{C},\, 
B \big) = 0$ in our theory which is equivalent to the symmetry transformations 
(10). We are entitled to make the above assertion because the transformations
(51) are equivalent and we have to make an appropriate choice of the transformations 
for our specific requirement.

We end this section with the remark that the trivial bosonic symmetry 
transformations $s^{(1)}_w\,\big(R,\,C,\, \bar{C},\, B \big) = 0$ lead 
to the derivation of the following vanishing basic brackets:
\begin{eqnarray}
&& [s,s^\dagger] \;= \;[s,l] \quad = \;[s, l^\dagger] \;\; 
=\; [s^\dagger, l] \;\;= \;[s^\dagger, l^\dagger] = 0, \nonumber\\
&& [c,s] \;\;\,= \;[c, s^\dagger] \;\;= \;[c,l] \quad 
=\; [c, l^\dagger] \;\;= \; [l, l^\dagger] = 0, \nonumber\\
&& [c^\dagger, s] \;\,=\; [c^\dagger, s^\dagger] \,
= \;[c^\dagger, l] \;\;=\; [c^\dagger, l^\dagger]\; = 0, \nonumber\\
&& [\bar{c}, s] \;\;\;= \; [\bar{c}, s^\dagger] \;\, 
= \;[\bar{c}, l] \quad =\; [\bar{c}, l^\dagger]\;\;\, = \;0, \nonumber\\
&& [\bar{c}^\dagger, s] \;\;= \; [\bar{c}^\dagger, s^\dagger]  
= \;[\bar{c}^\dagger, l] \;\;= \;[\bar{c}^\dagger, l^\dagger] \;\; = \; 0,
\end{eqnarray}
which are in complete agreement with the canonical basic brackets 
(cf. App. A), derived in Sec. 3. 
In a nut-shell, we draw the conclusion that {\it all} the \textit{six} 
continuous symmetries of our present theory lead to the derivation of 
basic canonical brackets that are in total agreement with the basic brackets 
derived by the standard canonical method of quantization.

\noindent
\section{Conclusions}

In our present endeavor, we have provided an alternative to the 
{\it standard} canonical method of quantization for a specific model 
of the Hodge theory which is nothing but the 1D rigid rotor. We have not used the
definition of canonical conjugate momenta w.r.t. 
the dynamical variables at any place in our approach which has led to the derivation 
of canonical basic brackets at the level of creation and annihilation operators of 
this theory. Our method of quantization depends heavily on the symmetry principles 
which provide an alternative to the definition of canonical conjugate 
momenta. However, we have taken the help of standard spin-statistics 
theorem in defining the (anti)commutators and utilized 
the concept of normal ordering to make sense out of the conserved Noether charges 
corresponding to the {\it six} continuous symmetries that are present in our theory.
As pointed out earlier, we would like to stress that, for the 1D system, the spin-statistics theorem 
is only limited to the definitions of (anti)commutators. There is no meaning of 
{\it spin} quantum number in 1D.

We would like to pin-point some of the subtle features of our present 
investigation. To obtain the normal mode expansion (18) for all the 
relevant variables, we have made physically motivated approximations 
where we have ignored the term [$({1/2})\,(r^2\,{\dot \theta}^2)$] from 
the Lagrangian (2) because it does {\it not} contribute anything in the
discussion of the continuous symmetries of our present theory.  
It has also been argued that, for a constant value of $\dot \theta$, this 
term becomes a constant in the case of a rigid rotor. As a consequence, 
we obtain the equations of motion:
$ \ddot{p_r} + p_r = 0$ and $ \ddot{\lambda} + \lambda = 0 $ which have very 
nice and simple normal mode expansion as illustrated in (18). We would 
like to add that, {\it even without} any approximation, we have the 
validity of the relationship: $ \frac{d^2}{dt^2} (\dot \lambda - p_r) 
+ (\dot \lambda - p_r) = 0$. One of the solutions of our interest 
(for this relationship) is $\ddot{\lambda} + \lambda = 0 $ and $ \ddot{p_r} 
+ p_r = 0$. These solutions are {\it not} unique but are of utmost importance 
to us as they support the normal mode expansions given in (18) for $\lambda(t)$ 
and $p_r(t)$ which are very useful to us.

We have applied our idea of quantization scheme to the discussion of 2D 
{\it free} Abelian gauge theory which is a model for the Hodge theory 
(see, e.g. [7]). It was interesting to extend this work to the case if 
{\it interacting} $U(1)$ gauge theory (i.e. QED) where the 1-form gauge 
field couples to the Dirac fields [8]. It has been very gratifying to 
observe that our method of quantization is true in the case of SUSY 
quantum mechanics where a SUSY harmonic oscillator is considered 
for its quantization [11]. We conjecture that our method of quantization 
would be valid for {\it all} the models for Hodge theory that would incorporate 
gauge theories, 1D toy models and SUSY theories. Having applied this method in the
context of gauge theories and SUSY theories, it was a challenging problem for 
us to apply it to a 1D toy model. We have accomplished this goal in our 
present investigation for the case of a 1D rigid rotor which happens to be 
a toy model for the Hodge theory [6].

Our method of quantization is valid {\it only} for a specific class of 
theories which are the models for the Hodge theory. For instance,
such theories are Abelian {\it p}-form ({\it p} 
= 1, 2, 3) gauge theories which have been shown to be the field theoretic 
models for the Hodge theory in $D = 2p$ dimensions of spacetime (see, e.g. 
[12-16]). These theories respect {\it six} continuous symmetries that lead to the 
derivation of canonical basic brackets amongst the
creation and annihilation operators. 
The (non-)vanishing brackets are exactly same as the ones derived by the 
{\it standard} method of canonical quantization scheme. 
Of course, our method is algebraically more involved but it has aesthetic 
appeal in the sense that it is the symmetry principles
that replace the definition of the canonical conjugate momenta. 
It is worth pointing out that, in a recent paper [11], we have applied our 
method of quantization to the supersymmetric (SUSY) ${\mathcal N} = 2$ 
harmonic oscillator and obtained the basic brackets from the symmetry principles. 
In this case, there are only {\it three} continuous symmetries
and they lead to the derivation of precise (anti)commutators that are also 
obtained by the standard canonical method.

We have proposed many models for the Hodge theory which are from the 
domains of $p$-form ($p = 1, 2, 3$) gauge theories 
[12-16] and ${\mathcal N } = 2$ SUSY quantum mechanics [17-19]. 
One of the decisive features of the models for the Hodge theories, 
connected with the $p$-form gauge theories, is that these theories 
are always endowed with {\it six} continuous symmetries within the 
framework of BRST formalism. On the contrary, all the models of 
${\mathcal N} = 2$ SUSY quantum mechanics 
(that have been shown to be the physical examples of Hodge theory 
[17-19]) respect only {\it three} continuous symmetries. We have 
established in [11] that these {\it three} symmetries 
are good enough to yield the proper (anti)commutators which are 
found to be exactly same as the {\it ones } derived by the {\it standard}  
canonical quantization method. It would be worthwhile to point out that,
for this purpose, the  
well-known one (0 + 1)-dimensional (1D) model of SUSY harmonic 
oscillator has been taken into consideration.
There is yet another SUSY quantum mechanical 
model which has been shown to be the physical example for the Hodge theory 
[20] where, once again, only {\it three} continuous symmetries exist. 
This is the simple toy model of ${\mathcal N} = 2 $ SUSY free particle. 
We plan to discuss its standard canonical quantization and wish to 
compare it with the quantization through symmetry principles.  It would 
be nice future endeavour for us to obtain the quantization of the above models 
[17 -19] by using our proposed {\it novel} method so that this idea could be 
firmly established [21]. \\

\noindent
{\bf Acknowledgements:} DS thanks UGC,
Government of India, New Delhi, for financial support through RFSMS scheme and 
TB is grateful to BHU-fellowship under which 
the present investigation has been carried out. \\

\newpage

\noindent
\begin{center}
\large{ \bf Appendix A: Brackets from the Canonical Method}
\end{center}
\vskip .9cm
We list here {\it all} the basic brackets (i.e. (anti)commutators) 
that emerge from the standard canonical method of quantization. 
As is evident from the main body of our text, the non-vanishing
canonical basic brackets are: $[\lambda, \, B] = i, \, [R,\, p_r] 
= i, \, \lbrace C, \, \dot{\bar{C}} \rbrace = +\,1 $
and $\lbrace \bar{C}, \,\dot{C}\rbrace = -\,1$ and the rest of the
brackets are zero. The trivial vanishing brackets are {\it eight}
in number (i.e. $[R,\,R] = [p_r,\, p_r] = [\lambda,\, \lambda] = [B, \, B] 
= \lbrace  C, \,C\rbrace = \lbrace \bar{C}, \, \bar{C} \rbrace = \lbrace \dot{C},
\, \dot{C}\rbrace = \lbrace \dot{\bar{C}}, \, \dot{\bar{C}}\rbrace = 0$) and
the rest of the basic brackets, that are equal to zero, are:
\begin{eqnarray}
&&[\lambda,\, C] \;\;\,= \, [\lambda,\, \bar{C}] \;\;= \,[\lambda,\, \dot{C}] \,\;
= \,[\lambda,\, \dot{\bar{C}}] \;= \,[\lambda,\, R] \quad = \,[\lambda, \, p_r] \,= 0,\nonumber\\
&&[B,\, C] \;\,=  \,[B,\, \bar{C}] \,\, = \,[B,\, \dot{C}] \,= \,[B,\, \dot{\bar{C}}] \,
= \,[B,\, R] \;\;\, = \,[B,\, p_r] = 0,\nonumber\\
&&\lbrace C, \, \bar{C} \rbrace = \, \lbrace C, \, \dot{C} \rbrace = \,[R, \, C] \,
= \,[p_r, \,C]  \,= \, \lbrace \dot{C}, \,  \dot{\bar{C}}\rbrace \, = \,[R, \, \dot{C}]\, =  0,\nonumber\\
&& \lbrace \bar{C} ,\, \dot{\bar{C}} \rbrace = \,[R, \, \bar{C}] \; = \,[p_r,\, \bar{C}] \,
= \,[R, \, \dot{\bar{C}}] \,= \,[p_r,\,\dot{C}] \;\;= \,[p_r, \, \dot{\bar{C}}] \,= 0.
\end{eqnarray}
Thus, the total number of (non-)vanishing brackets at the level of variables 
and their conjugate momenta are \textit{thirty six} in number. We shall express 
these in terms of the creation and annihilation operators of our present theory.

It is straightforward to note that the substitution of the mode expansions, 
in the above {\it thirty six} basic brackets, leads to the following sixty 
eight (68) vanishing basic (anti)commutators in terms of the creation 
and annihilation operators:  
\begin{eqnarray}
&&[s,\,s] \;\;= \;[s^\dagger,\,s^\dagger] \,\;= \;[s,\,k] \;\,
= \;[s^\dagger,\,k ^\dagger] = \; [s,\,d] \;\;= \;[s,\,d^\dagger] \;=  0, \nonumber\\
&&[s^\dagger,\, d] \,= \;[s^\dagger,\,d^\dagger] \;= \;[s,\,l] \;\;\,= \;[s, 
\, l^\dagger] \;\,= \;[s^\dagger,\, l] \;=\; [s^\dagger,\,l^\dagger]  =  0,\nonumber\\
&&[s,\,c] \; \;= \; [s, \,c^\dagger] \;\;\,=\; [s,\,\bar{c}] \;\;\;
= \;[s,\,\bar{c}^\dagger] \;\,= \;[s^\dagger,\,c] \;= \;[s^\dagger,\,c^\dagger] =  0, \nonumber\\
&&[s^\dagger,\,\bar{c}] \,= \;[s^\dagger,\,\bar{c}^\dagger] \;\,= \;[k,\,k] \;\,
= \;[k^\dagger, \,k^\dagger] = \;[k,\,d] \;\,= \;[k, \, d^\dagger] \;=  0, \nonumber\\
&& [k^\dagger, \, d] = \; [k^\dagger,\, d^\dagger] \,= \; [k,\,l] \;\;\,
= \; [k,\, l^\dagger] \;\,= \;[k^\dagger,\, l]\; = \;[k^\dagger,\,l^\dagger] = 0, \nonumber\\
&&[k,\,c] \;\,= \;[k,\,c^\dagger] \;\;= \;[k,\,\bar{c}] \;\;\,= \;[k,\,\bar{c}^\dagger] \;\,
= \;[k^\dagger,\,c] \;= \;[k^\dagger,\,c^\dagger] =  0,\nonumber\\
&&[k^\dagger,\,\bar{c}] \,= \;[k^\dagger,\,\bar{c}^\dagger] \,
= \; [d,\,d] \;\;= \;[d^\dagger,\,d^\dagger] \, = \;[d,\,l] \;\;
= \;[d^\dagger,\,l^\dagger] \,= 0, \nonumber\\
&&[d,\,c] \;\;= \;[d,\,c^\dagger] \;\;= \;[d,\,\bar{c}] \;\;\,
= \;[d,\,\bar{c}^\dagger] \;\;= \;[d^\dagger,\,c] \;= \;[d^\dagger,\,c^\dagger] =   0, \nonumber\\
&& [d^\dagger,\,\bar{c}] \,= \;[d^\dagger,\,\bar{c}^\dagger] \;= \;[l,\,l] \;\;\,
= \;[l^\dagger,\,l^\dagger] \;\,=
 \;[l,\,c] \;\;= \;[l,\,c^\dagger] \;\;= 0, \nonumber\\
&&[l,\,\bar{c}]\;\; = \;[l,\,\bar{c}^\dagger] \;\;\,= \;[l^\dagger,\,c] \;\,
= \;[l^\dagger,\,c^\dagger] \;\,= \;[l^\dagger,\,\bar{c}] \;
= \;[l^\dagger,\,\bar{c}^\dagger]  \,= 0, \nonumber\\
&&\lbrace c,\, c\rbrace \,= \;\lbrace c,\, c^\dagger \rbrace \,
= \,\lbrace c^\dagger,\, c^\dagger \rbrace = \,\lbrace c, \bar{c} \rbrace \;\;\;
 = \,\lbrace \bar{c},\, \bar{c} \rbrace \;= \;\lbrace \bar{c},\, 
 \bar{c}^\dagger \rbrace = 0, \nonumber\\
&&\lbrace \bar{c}^\dagger,\, \bar{c}^\dagger \rbrace 
= \lbrace c^\dagger,\, \bar{c}^\dagger \rbrace = 0, \nonumber\\
\end{eqnarray}
along with the following non-vanishing canonical basic brackets 
\begin{eqnarray}
&&[d,\,l^\dagger] = [d^\dagger,\,l] = i, \quad [s,\,k^\dagger] = [s^\dagger,\, k] 
= i, \quad \lbrace c,\,\bar{c}^\dagger \rbrace = - i, \; \lbrace 
\bar{c},\,c^\dagger \rbrace = +\,i.
\end{eqnarray}
This exercise has been performed so that we can derive all these brackets 
from the symmetry principles and compare them in a precise manner. 
The salient features of the above basic brackets are as follows. 
First, we note that  there are only {\it four} independent brackets 
that are non-vanishing [cf. (60)]. We point out that the brackets
 $[d, \, l^\dagger] = i$ and $[d^\dagger,\, l] = i$ are Hermitian 
 conjugate of each-other. Thus, only one of them  is independent. 
 Second, all the brackets in (59) are {\it not} independent (for instance, 
 $[s^{\dagger},\, c] = 0 $ is equivalent to $[s,\,c^\dagger] = 0$ because 
 these are Hermitian conjugate of each-other). Third, the basic brackets
$\{ c, \,\bar c^\dagger \} = - \,i $ and $\{ \bar c,\, c^\dagger \} = + \,i $ 
are independent of each-other because the variables $ C(t)$ and $\bar C(t)$ 
have been taken to be independent right from the beginning. Finally, if 
one of the bracket is calculated from the symmetry principles, its Hermitian 
conjugate would {\it also} be automatically true. This input has been taken 
into account in the main body of our text.

\vskip .9cm
\noindent
\begin{center}
\large{ \bf Appendix B: Logical Approximations and Mode Expansions}
\end{center}
\vskip .9cm
Here we discuss some of the details of our approximation as well as solution of the 
equations of motion: $B = - \,(\dot{\lambda} - p_r), \; B = d/dt\,(r - a), \; \dot{B} 
= -\,(r - a), \;\dot{p_r} + \lambda = 0$ which emerge from the approximated Lagrangian 
$L^{(0)}_B = \dot{r}\,p_r - \lambda\, (r - a) + B\,(\dot{\lambda} - p_r) + B^2/2 - i\,
 \dot{\bar{C}}\, \dot{C} + i\, \bar{C}\, C$. In the latter, one term (i.e. $(r^2\,
\dot{\theta}^2)/2$) has been ignored for a rigid rotor with a constant angular
 velocity $\dot{\theta}$ (i.e. $\dot{\theta}$ = constant). The above EL equations
 of motion imply that $\ddot{B} + B = 0,\; \ddot{R} + R = 0$ and $\frac{d^2}{dt^2}\,
(\dot{\lambda} - p_r) + (\dot{\lambda} - p_r) = 0$ where $R = (r - a)$. 
Using the equation of motion $\dot{p_r} + \lambda  = 0$, one can clearly observe that
the following is true, namely;
\begin{eqnarray}
\frac{d^2}{dt^2}\,(\dot{\lambda} - p_r) + (\dot{\lambda} - p_r) = 0 
\quad \Longrightarrow \quad \frac{d^2}{dt^2}\,(\ddot{p_r} + p_r) + (\ddot{p_r} + p_r) 
= 0, 
\end{eqnarray}
which is a common feature of an equation of motion for a harmonic oscillator 
in terms of the variable $p_r\,(t)$( i.e. $\ddot{p_r} + \omega^2\, p_r = 0$ 
with frequency $\omega = 1$). It is evident that $\frac{d^2}{dt^2}\,(\ddot{p_r} + p_r) 
+ (\ddot{p_r} + p_r) = 0$ would be satisfied if we set $\ddot{p_r} + p_r = 0$. 
It may be worthwhile
 to mention that all equations like $\frac{d^{2n}}{dt^{2n}}\,(\ddot{p_r} + p_r) + 
\frac{d^{(2n-2)}}{dt^{(2n-2)}}\,(\ddot{p_r} + p_r) = 0 \;\;(n= 1, 2, 3...)$ would
 be always satisfied for the EOM connected to the harmonic oscillator 
$\ddot{p_r} + p_r = 0$ with frequency  $ \omega = 1 $.
The solution of (61) is the one which is given in the mode expansion (18). 
The equation of motion $\dot{p_r} + \lambda = 0$ implies that one of the 
interesting solutions of this equation of motion that could be satisfied by $\lambda$ 
would be $\ddot{\lambda} + \lambda = 0$ whose mode expansion is given in (18).

We would like to observe that the normal mode expansion (18) have been
taken in a uniform manner because $\ddot{C} + C = 0,\, \ddot{\bar{C}} 
+ \bar{C} = 0,\; \ddot{R} + R = 0,\; \ddot{B} + B = 0$ emerge 
automatically but $\ddot{p_r} + p_r = 0$ and $\ddot{\lambda} 
+ \lambda = 0$ come out due to some approximations. 
It can be checked that if we take the constraint equations:  
$R \approx 0$ and $d/dt\,(R)\approx  0$, the equations of motion 
$\ddot{\lambda} + \lambda = 0 $ and $\ddot{p_r} + p_r = 0$ 
emerge very naturally from our theory. We lay emphasis on the fact that
the equations of motion (i.e.  $\ddot{p_r} + p_r = 0$ and $\ddot{\lambda} + \lambda 
= 0 $) for $p_r (t)$ and $\lambda (t)$ are due to specific approximations but these 
are the ones which are interesting for our purposes.

\end{document}